\documentclass{llncs}
\usepackage{llncsdoc}

\usepackage{epsfig}
\usepackage{graphicx}
\usepackage{tabularx}
\usepackage{multirow}
\usepackage{color}
\usepackage{subfigure}

\begin{document}

\title{Analyzing Tag Distributions in Folksonomies for Resource Classification}

\author{Arkaitz Zubiaga \and Raquel Mart\'{i}nez \and V\'{i}ctor Fresno}
\institute{NLP \& IR Group @ UNED}
% \email{\{azubiaga,vfresno,raquel\}@lsi.uned.es}

\maketitle

\begin{abstract}
Recent research has shown the usefulness of social tags as a data source to feed resource classification. Little is known about the effect of settings on folksonomies created on social tagging systems. In this work, we consider the settings of social tagging systems to further understand tag distributions in folksonomies. We analyze in depth the tag distributions on three large-scale social tagging datasets, and analyze the effect on a resource classification task. To this end, we study the appropriateness of applying weighting schemes based on the well-known TF-IDF for resource classification. We show the great importance of settings as to altering tag distributions. Among those settings, tag suggestions produce very different folksonomies, which condition the success of the employed weighting schemes. Our findings and analyses are relevant for researchers studying tag-based resource classification, user behavior in social networks, the structure of folksonomies and tag distributions, as well as for developers of social tagging systems in search of an appropriate setting.
\end{abstract}

\section{Introduction}
\label{introduction}

Social tagging systems allow users to annotate resources with tags in an aggregated way. These systems generate large amounts of metadata that increase the availability of resource descriptions, and can enhance information access in a wide sense \cite{heymann_can_social_2008,awawdeh2010ksem}. It has attracted a large number of researchers to using them for improving resource retrieval, organization and classification tasks, among others \cite{gupta2010survey}. Regarding resource classification tasks, social tags have shown high effectiveness by outperforming content-based approaches \cite{zubiaga_getting_2009,noll_metadata_2008,godoy2010exploiting}. Social tags represent an interesting data source for the sake of resource classification tasks.

However, each social tagging system may produce a different folksonomy structure. So far, research on tag-based classification has focused on analyzing the use of social tags for specific datasets, and no attention has been paid to settings of different systems \cite{zubiaga_getting_2009,godoy2010exploiting}. We aim at complementing earlier research by further analyzing tag distributions as a way of weighting the relevance of tags in the collection. As an approach to considering distributions as an indicator of the relevance of tags in a collection, we perform a classification study by adapting the TF-IDF weighting scheme \cite{salton88termweighting}. Text document collections are simply made up by documents containing terms, whereas social tagging systems involve different users and bookmarks for each resource containing tags. Several works have applied the TF-IDF scheme to social tagging systems \cite{angelova2008,shepitsen2008,li2008,liang2010}, but there is no evidence on its suitability for these systems.

In this paper, we analyze the settings, and study folksonomies and tag distributions on three large-scale social tagging datasets. We apply three weighting schemes based on TF-IDF to define the relevance of tags on a folksonomy, and explore their suitability according to system's settings. We evaluate the results by performing resource classification experiments, comparing those weighting schemes to the sole use of TF in the absence of weights. We find that tag distributions in folksonomies can help determine relevance of tags, but they are strictly subject to settings of the tagging system. Specifically, an IDF-like weighting scheme does not present the desired effect when a social tagging system suggests tags to the user. where utterly different folksonomy structures are produced.

The paper is organized as follows. Next, we define a collaborative tagging system in Section \ref{collaborative-tagging}. We detail the studied collaborative tagging systems, the process of generation of the datasets, and perform a thorough analysis of tag distributions in Section \ref{datasets}. We provide a brief overview of TF-IDF, and introduce the variants adapted to collaborative tagging in Section \ref{tag-weighting-functions}. Then, we present the tag-based resource classification experiments and analyze the results in Section \ref{tag-based-classification}. We conclude the paper with our thoughts in Section \ref{conclusions}.

\section{Social Tagging}
\label{collaborative-tagging}

Tagging is an open way that allows users to bookmark and annotate with tags their favorite resources (e.g., web pages, movies or books). Tagging resources facilitates future retrieval by relying on tags as metadata describing resources. On social tagging systems, users can collaboratively annotate resources, so that many users can tag the same resource. Tags by different users aggregated on a resource provide vast amounts of metadata. The collection of tags defined by users of a system creates a tag-based organization, so-called folksonomy. The subset of tags assigned by a single user creates a smaller folksonomy, also known as personomy. For instance, CiteULike\footnote{http://www.citeulike.org}, LibraryThing and Delicious are social tagging system where users collaboratively annotate resource. On these systems, each resource (papers, books and URLs, respectively) can be annotated and tagged by all the users who consider it interesting. On a social tagging system, there is a set of users ($U$), who are posting bookmarks ($B$) for resources ($R$) annotated by tags ($T$). Each user $u_{i} \in U$ can post a bookmark $b_{ij} \in B$ of a resource $r_{j} \in R$ with a set of tags $T_{ij} = \{t_{ij1},...,t_{ijp}\}$, with a variable number $p$ of tags. After $k$ users posted $r_{j}$, it is described with a weighted set of tags $T_{j} = \{w_{j1} t_{j1},...,w_{jn} t_{jn}\}$, where $w_{j1},...,w_{jn} \leq k$ represent the number of assignments of a specific tag. Accordingly, each bookmark is a triple of a user, a resource, and a set of tags: $b_{ij}: u_{i} \times r_{j} \times T_{ij}$. Thus, each user saves bookmarks of different resources, and a resource has bookmarks posted by different users. The result of aggregating tags within bookmarks by a user is known as the personomy of the user: $T_{i} = \{w_{i1} t_{i1},...,w_{im} t_{im}\}$, where $m$ is the number of different tags in user's personomy.

For instance, a user could tag this work as \texttt{social-tagging}, \texttt{research}, and \texttt{paper}, whereas another user could use the tags \texttt{classification}, \texttt{paper}, \texttt{social-tagging}, and \texttt{social-bookmarking} to annotate it. The aggregation of these annotations would produce the following: \texttt{social-tagging} (2), \texttt{paper} (2), \texttt{social-bookmarking} (1), \texttt{classification} (1), and \texttt{research} (1). In this example the values represent the weighted union of all tags.

\section{Datasets}
\label{datasets}

Next, we introduce and analyze the three large-scale datasets we gathered from well-known social tagging sites: Delicious, LibraryThing, and GoodReads. All of them had been gathered from March to May 2010.

\subsection{Studied Social Tagging Sites}
\label{social-tagging-sites}

\textbf{Delicious} is a social tagging site that allows users to save and tag their preferred web pages, in order to ease the subsequent navigation and retrieval on large collections of annotated bookmarks. On a social bookmarking site, any web page can be saved, so that the range of covered topics can become as wide as the Web is. It is known that the site is biased to some computer and design related topics though. When a user saves a URL as a bookmark, the system suggests tags previously used for that URL if some users had annotated it before. Thus, new annotators can easily add tags used by earlier users without typing them.

\textbf{LibraryThing} and \textbf{GoodReads} are social tagging sites where users save and annotate books. Commonly, users annotate the books they own, they have read, or they are planning to read. Besides readers, there are also well-known writers and libraries contributing as users on these sites. The main difference among these two systems is that LibraryThing does not suggest tags when saving a book, whereas GoodReads lets the user select from tags within his personomy, that is, tags he previously assigned to other books. The latter makes it easier to reuse users' favorite tags, without re-typing them. Another difference is that LibraryThing allows some users to group tags with the same meaning, linking thus typos, synonyms and translations to a single tag, e.g., \textit{science-fiction}, \textit{sf} and \textit{ciencia ficci\'on} are grouped into \textit{science fiction}.

Despite of the aforementioned differences, all of them have some characteristics in common: users save resources as bookmarks, a bookmark can be annotated by a variable number of tags ranging from zero to unlimited, and the vocabulary of the tags is open. Table \ref{tab:st-features} summarizes the features of the social tagging sites we study in this work.

\begin{table}[htb]
\begin{center}
 \begin{tabular}{|p{2.7cm}|p{2.7cm}|p{2.7cm}|p{2.7cm}|}
  \hline
  & \small \textbf{Delicious} & \small \textbf{LibraryThing} & \small \textbf{GoodReads} \\
  \hline
  \small \textbf{Resources} & \small web documents & \small books & \small books \\
  \hline
  \small \textbf{Tag suggestions} & \small tags from earlier bookmarks of the resource & \small no & \small tags in user's personomy \\
  \hline
  \small \textbf{Users} & \small general & \small readers, writers \& libraries & \small readers, writers \& libraries \\
  \hline
  \small \textbf{Tag grouping} & \small no & \small selected users suggest merging tags & \small no \\
  \hline
  \small \textbf{Vocabulary} & \small open & \small open & \small open \\
  \hline
  \small \textbf{Tag insertion} & \small space-separated & \small comma-separated & \small one by one text-box \\
  \hline
 \end{tabular}
\end{center}
\caption{Characteristics of studied social tagging systems.}
\label{tab:st-features}
\end{table}

\subsection{Generation Process of Datasets}
\label{datasets-generation}

First of all, we queried the three sites for popular resources\footnote{We consider a resource to be popular if at least 100 users have annotated it as a bookmark. It was shown that the tag set of a resource tends to converge when that many users contribute to it \cite{golder_structure_2006}.}. In the case of Delicious, we gathered a set of 87,096 URLs. As regards to LibraryThing and GoodReads, we found an intersection of 65,929 popular books. In the next step, we looked for classification labels assigned by experts for both kind of resources. For Delicious, we used the Open Directory Project\footnote{http://www.dmoz.org} (ODP) as a ground truth, where we found an intersection of 12,616 URLs with our set. For the set of books, we fetched their classification for both the Dewey Decimal Classification (DDC) and the Library of Congress Classification (LCC) systems. The former is a classical taxonomy that is still widely used in libraries, whereas the latter is used by most research and academic libraries. We found that 27,299 books were categorized on DDC, and 24,861 books have an LCC category assigned.
% In total, there are 38,149 books with category data from either one or both category schemes. An important feature of these categorized resources is that all of them show a marked imbalancement in the distribution across the categories.

Finally, we queried (a) Delicious for all the public bookmarks from users involved in the set of categorized URLs, and (b) LibraryThing and GoodReads for all the public bookmarks from users involved in the set of categorized books. This resulted in a large collections of bookmarks, not only for categorized resources, but also for many others. Each bookmark comprises: (1) the user who annotated, (2) the annotated resource, and (3) the associated tags. We saved all the tags attached to each bookmark, except for GoodReads, where a tag is automatically attached to each bookmark depending on the reading state of the book: \textit{read}, \textit{currently-reading} or \textit{to-read}. We do not consider this to be part of the tagging process, but just an automated step, and we removed all their appearances in the dataset.

\subsection{Statistics and Analysis of the Datasets}
\label{dataset-stats}

It is worthwhile noting that attaching tags to a bookmark is an optional step, so that depending on the social tagging site, a number of bookmarks may remain without tags. Table \ref{tab:dataset-stats} presents the number of users, bookmarks and resources we gathered for each of the datasets, as well as the percent with attached annotations. In this work, as we rely on tagging data, we only consider annotated data, ruling out bookmarks without tags. Thus, from now on, all the results and statistics presented are based on annotated bookmarks. From these statistics, it stands out that most users (above 87\%) provide tags for bookmarks on Delicious, whereas there are fewer users who tend to assign tags to resources on LibraryThing and GoodReads (roughly 38\% and 17\%, respectively).

\begin{table}[htb]
\begin{center}
 \begin{tabular}{|l|c|c|c|}
  \hline
  \multicolumn{4}{|c|}{\small \textbf{Delicious}} \\
  \hline
  & \small Annotated & \small Total & \small Percent \\
  \hline
  \small Users & \small 1,618,635 & \small 1,855,792 & \small 87.22\% \\
  \hline
  \small Bookmarks & \small 273,478,137 & \small 300,571,231 & \small 91.00\% \\
  \hline
  \small Resources & \small 92,432,071 & \small 102,828,761 & \small 89.89\% \\
  \hline
  \multicolumn{2}{|l|}{\small Tags} & \small 11,541,977 & \small - \\
  \hline
  \multicolumn{4}{|c|}{\small \textbf{LibraryThing}} \\
  \hline
  & \small Annotated & \small Total & \small Percent \\
  \hline
  \small Users & \small 153,606 & \small 400,336 & \small 38.37\% \\
  \hline
  \small Bookmarks & \small 22,343,427 & \small 44,612,784 & \small 50.08\% \\
  \hline
  \small Resources & \small 3,776,320 & \small 5,002,790 & \small 75.48\% \\
  \hline
  \multicolumn{2}{|l|}{\small Tags} & \small 2,140,734 & \small - \\
  \hline
  \multicolumn{4}{|c|}{\small \textbf{GoodReads}} \\
  \hline
  & \small Annotated & \small Total & \small Percent \\
  \hline
  \small Users & \small 110,344 & \small 649,689 & \small 16.98\% \\
  \hline
  \small Bookmarks & \small 9,323,539 & \small 47,302,861 & \small 19.71\% \\
  \hline
  \small Resources & \small 1,101,067 & \small 1,890,443 & \small 58.24\% \\
  \hline
  \multicolumn{2}{|l|}{\small Tags} & \small 179,429 & \small - \\
  \hline
 \end{tabular}
\end{center}
\caption{Statistics on availability of tags in users, bookmarks, and resources for the three datasets.}
\label{tab:dataset-stats}
\end{table}

\begin{figure*}[htb]
\centering
\epsfig{file=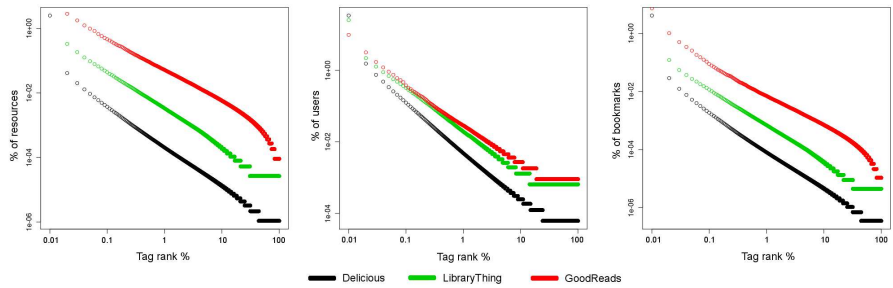, width=120mm}
\caption{Tag usage percentages. These 3 graphs represent, on a logarithmic scale for both \textit{x} and \textit{y} axes, the percent of annotations to resources, users, and bookmarks per tag rank on each site.}
\label{fig:tag-usage}
\end{figure*}

Regarding the distribution of tags across all the resources, users and bookmarks in the datasets, there is a clear difference of behavior among the three collections. Figure \ref{fig:tag-usage} shows the percent of resources, users and bookmarks on which tags are annotated according to their rank on the system. That is, the X axis refers to the percent of the tag rank, whereas the Y axis represents the percent of appearances in resources, users and bookmarks. For instance, if the tag ranked first had been annotated on the half of the resources, the value for the top ranked tag on resources would be 50\%. Thus, these graphs enable to analyze how popular are the tags in the top as compared to the tags in the tail on each site. It stands out that GoodReads has the highest usage of tags in the tail, but Delicious presents the highest usage of tags in the top. Delicious is the site with highest diversity of tags, where a few tags become really popular, and many tags are seldom-used (note the logarithmic scale of the graphs). We believe that the reasons for these differences on tag distributions are:

\begin{itemize}
 \item Since Delicious suggests tags that have been annotated by previous users to a resource, it is obvious that those tags on the top are likely to happen more frequently, whereas others may barely be used.
 \item LibraryThing and GoodReads do not suggest tags used by earlier users and, therefore, tags other than those on the top tend to be used more frequently than in Delicious.
 \item GoodReads suggests tags from previous bookmarks of the same user, instead of tags that others assigned to the resource being tagged. Thus, this encourages reusing tags in their personomy, making it remain with a smaller number of tags (see Table \ref{tab:tag-per-item}). In addition, users tend to assign fewer tags to a bookmark on average, probably due to the one-by-one tag insertion of site's interface.
\end{itemize}

\begin{table}[htb]
\begin{center}
 \begin{tabular}{|l|c|c|c|}
  \hline
  \textbf{\# of tags} & \small \textbf{Delicious} & \small \textbf{LibraryThing} & \small \textbf{GoodReads} \\
  \hline
  \small Per resource & \small 33.35 & \small 14.53 & \small 13.33 \\
  \hline
  \small Per user & \small 632.714 & \small 357.15 & \small 131.03 \\
  \hline
  \small Per bookmark & \small 3.75 & \small 2.46 & \small 1.55 \\
  \hline
 \end{tabular}
\end{center}
\caption{Average counts of different tags}
\label{tab:tag-per-item}
\end{table}

\begin{figure}
    \centering
    \subfigure[Tag distribution across resources (r), users (u) and bookmarks (b).]
    {
        \includegraphics[width=55mm]{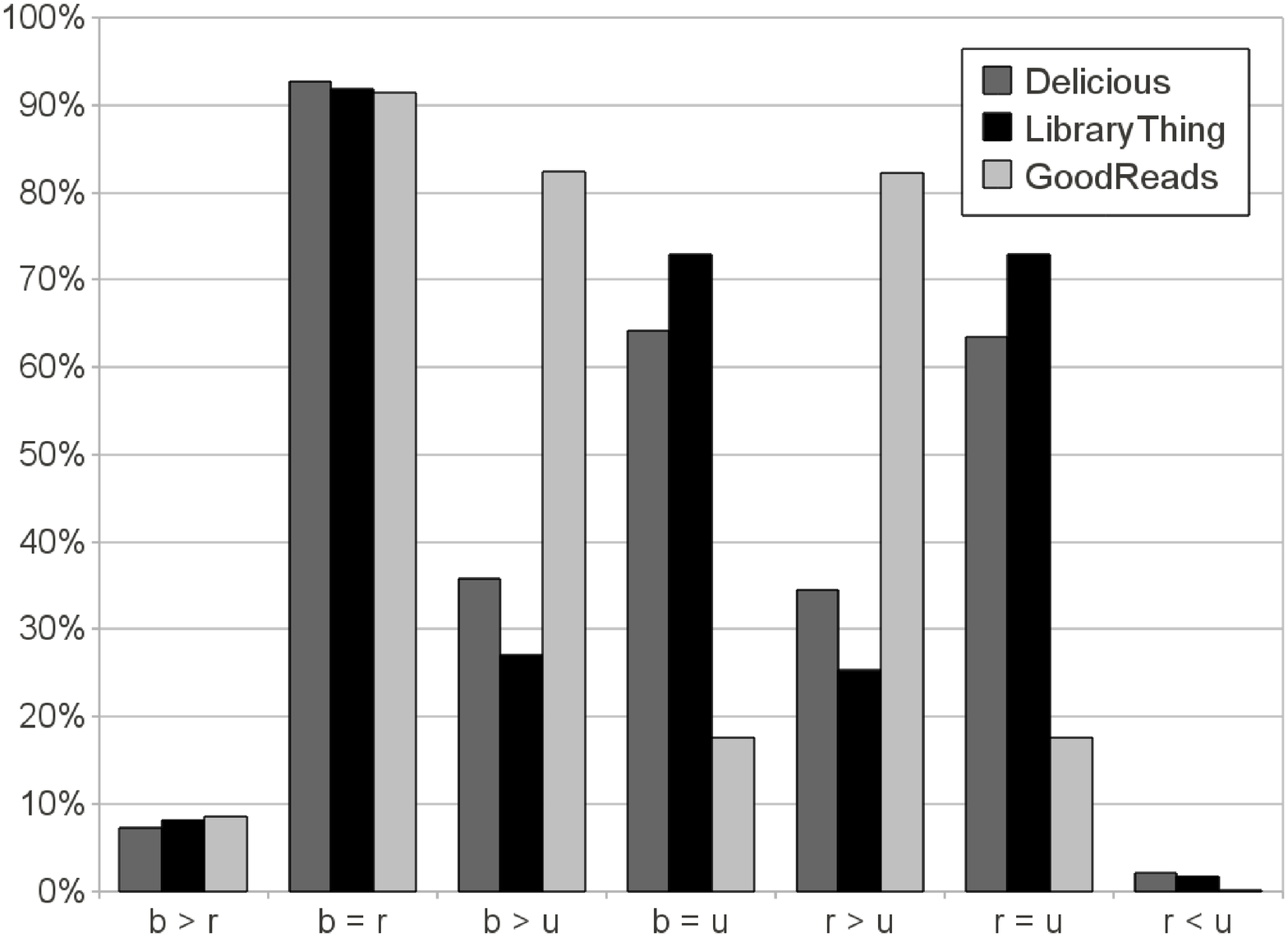}
        \label{fig:tag-distribution}
    }
    \subfigure[Novelty ratio of tags per rank of bookmark.]
    {
        \includegraphics[width=55mm]{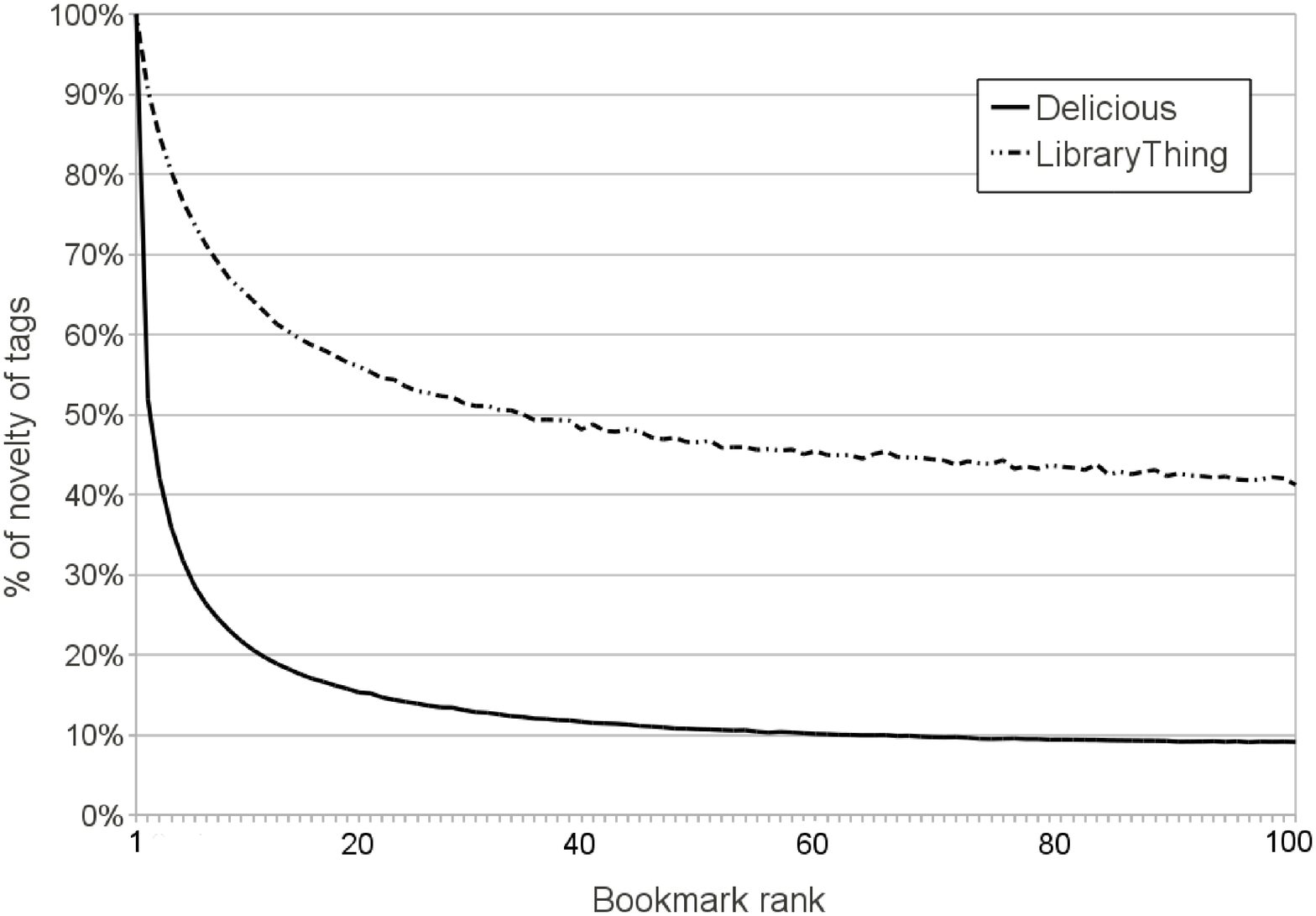}
        \label{fig:tag-novelty}
    }
    \label{fig:sample_subfigures}
\end{figure}

Regarding the distribution of tags across resources, users, and bookmarks, Figure \ref{fig:tag-distribution} shows percents of tags appearing more, equal or less frequently in an item (i.e., resources, users or bookmarks) than in another. It is obvious that a tag cannot appear in smaller number of bookmarks than users or resources, by definition. Looking at the rest of data, it stands out that tags tend to appear in more bookmarks than users ($b > u$) and more resources than users ($r > u$) for GoodReads, due to the same feature that allows users to select among tags in their personomy. However, LibraryThing and Delicious have many tags present in the same number of bookmarks and users ($b = u$), and resources and users ($r = u$), even though the difference is more marked for the former site. This reflects the large number of tags that users utilize just once on these sites. All three sites have two features in common: there are a few exceptions of tags utilized by more users than the number of resources it appears in ($r < u$), and almost all the tags are present in the same number of bookmarks and resources ($b = r$). The latter, combined with the lower ($b = u$) values, means that there is a large number of users spreading personal tags across resources that only have a bookmark with that tag, especially on GoodReads, but also for the other two sites.

Finally, we analyze the evolution of annotations throughout the time, i.e, the extent to which a bookmark introduces new tags into a resource that were not present in earlier bookmarks. Figure \ref{fig:tag-novelty} shows these statistics for Delicious and LibraryThing. The plot for GoodReads is not shown because neither the timestamp nor the ordering of the bookmarks is available in our dataset. The graph shows, on average, the ratio of new tags, not present in earlier bookmarks of a resource, assigned in bookmarks that rank from first to 100th bookmark, i.e., if $tag_1$ and $tag_2$ were annotated in the first bookmark of a resource, and $tag_2$ and $tag_3$ in the second bookmark for the same resource, the ratio of novelty for the second bookmark is of 50\%. It stands out the marked inferiority of tag novelty on Delicious as against to LibraryThing. This is, again, due to the tag suggestion policy of Delicious, what makes previous tags to re-occur more frequently.

\section{Tag Weighting Functions}
\label{tag-weighting-functions}

\subsection{TF-IDF as a Term Weighting Function}
\label{tfidf}

TF-IDF is a term weighting function that combines the term frequency and inverse document frequency to produce a composite weight for each term in a document \cite{salton88termweighting}. This weight is higher when $t$ occurs many times within a small number of documents (thus contributing to high discriminating power to those documents); lower when the term occurs fewer times in a document, or occurs in many documents (thus offering a less pronounced relevance signal); and it becomes null when the term occurs in all the documents.

TF-IDF is the most widely used weighting function for representing documents in a vector-space model, and has become a ``de facto standard'' for automated classification tasks \cite{joachims_text_1998}.

\subsection{TF-IDF for Tags}
\label{tfidf-for-tags}

In order to analyze the effect of tag distributions on each social tagging system to the resource classification task, we adapt the classical TF-IDF function. Unlike classical collections of web documents or library catalogs, where the terms distribute across documents in the collection, social tagging systems provide further data as new dimensions to explore into. Besides tags' distribution across documents or annotated resources, different users set those tags within different bookmarks. These two characteristics are specific to social tagging, and were not available on classical text collections.

Next, we introduce three tag weighting approaches, taking the classical TF-IDF approach to the social tagging scenario, and adapting it to rely on resources, users and bookmarks. These three dimensions suggest defining that many tag weighting functions considering inverse resource frequency (IRF), inverse user frequency (IUF), and inverse bookmark frequency (IBF) values, respectively. The three approaches we compare in this work follow the same function for the tag \textit{i} within the resource \textit{j}:

$$TF\mbox{-}IxF_{ij} = tf_{ij} \cdot ixf$$

where $tf_{ij}$ is the number of occurrences of the tag \textit{i} in the resource \textit{j}, and $ixf$ is the inverse frequency function considered in each case, $irf$, $iuf$ or $ibf$, thus $x$ being $r$, $u$, or $b$. Accordingly, we define TF-IRF for resources, TF-IUF for users, and TF-IBF for bookmarks.

There is little work using TF-IRF, usually referred to as TF-IDF. \cite{angelova2008} rely on this measure to infer similarity of tags by creating a tag graph, weighting the TF-IDF value of each user to a tag. \cite{shepitsen2008} and \cite{liang2010} use this measure to represent the resources in a recommendation system where resources are recommended to users. \cite{li2008} create tag vectors using TF-IDF to compute the similarity between two documents annotated on Delicious.

TF-IUF was inferred from a previous application to a collaborative filtering system \cite{breese98empiricalanalysis}. With the aim of recommending resources to users, \cite{diederich2006} and \cite{liang2010} rely on the IUF for discovering similarities among users. The latter use both IUF and IRF to represent users and resources, respectively, but no comparison is performed among their characteristics. In \cite{abbasi2009}, TF-IUF is used along with TF-IRF over Flickr for finding landmark photos.

To the best of our knowledge, TF-IBF has never been used so far. Even though all three frequencies can somehow be related, there are substantial differences among them. A tag used by many users can spread across many resources, or it can just congregate in a few resources. Likewise, this factor might affect the number of bookmarks.

\section{Tag-based Classification}
\label{tag-based-classification}

Next, we present the classification experiments that enable (1) the analysis of how each of the tag weighting functions contributes to the classification of annotated resources and whether they outperform the baseline relying only on the tag frequency (TF), as well as (2) discovering whether the settings of social tagging systems affect the performance of the tag weighting functions.

In order to analyze the impact of weighting functions on tag-based classification, we use multiclass Support Vector Machines (SVM) \cite{joachims_text_1998}. Specifically, we use ''svm-multiclass''\footnote{http://svmlight.joachims.org/svm\_multiclass.html}, with the linear kernel and the default parameters.

We perform the classification tasks relying on the top level of the taxonomies. We maintained the structure of all the taxonomies, but merged the categories E (\textit{History of America}) and F (\textit{History of the United States and British, Dutch, French, and Latin America}) on LCC, as the differences between them do not seem clear. Thus, ODP is composed by 17 categories, DDC by 10, and LCC by 20. We used different training set sizes, and made 6 different selections for each size, getting the average of all 6 runs not to make the results depend on the selected training instances.

\begin{table}[htb]
 \begin{center}
  \begin{tabular}{ l @{}r@{} @{}r@{} @{}r@{} @{}r@{} @{}r@{} @{}r@{} @{}r@{} }
   \hline
   \vspace{-12pt} \\
   \multicolumn{8}{ c }{\small \textbf{Delicious - ODP}} \\
   \vspace{-12pt} \\
   \hline
   \small training set\textcolor{white}{\_} & \small \textcolor{white}{\_}600{\hspace{1px}} & \small {\hspace{1px}}1400{\hspace{1px}} & \small {\hspace{1px}}2200{\hspace{1px}} & \small {\hspace{1px}}3000{\hspace{1px}} & \small {\hspace{1px}}4000{\hspace{1px}} & \small {\hspace{1px}}5000{\hspace{1px}} & \small {\hspace{1px}}6000{\hspace{1px}} \\
   \hline
   \vspace{-12pt} \\
   \small TF & \small \textbf{.533} & \small \textbf{.600} & \small \textbf{.629} & \small \textbf{.647} & \small \textbf{.660} & \small \textbf{.669} & \small \textbf{.680} \\
   \small TF-IRF & \small .516 & \small .571 & \small .593 & \small .607 & \small .619 & \small .631 & \small .639 \\
   \small TF-IBF & \small .519 & \small .573 & \small .596 & \small .611 & \small .622 & \small .633 & \small .641 \\
   \small TF-IUF & \small .528 & \small .580 & \small .607 & \small .625 & \small .636 & \small .653 & \small .661 \\
   \hline
   \vspace{-12pt} \\
   \multicolumn{8}{ c }{\small \textbf{LibraryThing - DDC}} \\
   \vspace{-12pt} \\
   \hline
   \small training set\textcolor{white}{\_} & \small {\hspace{1px}}3000{\hspace{1px}} & \small {\hspace{1px}}6000{\hspace{1px}} & \small {\hspace{1px}}9000{\hspace{1px}} & \small {\hspace{1px}}12000{\hspace{1px}} & \small {\hspace{1px}}15000{\hspace{1px}} & \small {\hspace{1px}}18000{\hspace{1px}} & \small {\hspace{1px}}21000{\hspace{1px}} \\
   \hline
   \vspace{-12pt} \\
   \small TF & \small .861 & \small .864 & \small .864 & \small .867 & \small .869 & \small .869 & \small .868 \\
   \small TF-IRF & \small .877 & \small .889 & \small .894 & \small \textbf{.897} & \small \textbf{.900} & \small .902 & \small .902 \\
   \small TF-IBF & \small .877 & \small .889 & \small .894 & \small \textbf{.897} & \small \textbf{.900} & \small \textbf{.903} & \small \textbf{.904} \\
   \small TF-IUF & \small \textbf{.881} & \small \textbf{.891} & \small \textbf{.895} & \small \textbf{.897} & \small .899 & \small .901 & \small .900 \\
   \vspace{-12pt} \\
   \hline
   \vspace{-12pt} \\
   \multicolumn{8}{ c }{\small \textbf{LibraryThing - LCC}} \\
   \vspace{-12pt} \\
   \hline
   \small training set\textcolor{white}{\_} & \small {\hspace{1px}}3000{\hspace{1px}} & \small {\hspace{1px}}6000{\hspace{1px}} & \small {\hspace{1px}}9000{\hspace{1px}} & \small {\hspace{1px}}12000{\hspace{1px}} & \small {\hspace{1px}}15000{\hspace{1px}} & \small {\hspace{1px}}18000{\hspace{1px}} & \small {\hspace{1px}}21000{\hspace{1px}} \\
   \hline
   \vspace{-12pt} \\
   \small TF & \small .853 & \small .857 & \small .856 & \small .861 & \small .861 & \small .857 & \small .861 \\
   \small TF-IRF & \small .867 & \small \textbf{.883} & \small .887 & \small \textbf{.893} & \small .895 & \small .894 & \small .897 \\
   \small TF-IBF & \small .867 & \small \textbf{.883} & \small \textbf{.888} & \small \textbf{.893} & \small \textbf{.896} & \small \textbf{.895} & \small \textbf{.898} \\
   \small TF-IUF & \small \textbf{.871} & \small .882 & \small .885 & \small .892 & \small .893 & \small .892 & \small .894 \\
   \hline
   \vspace{-12pt} \\
   \multicolumn{8}{ c }{\small \textbf{GoodReads - DDC}} \\
   \vspace{-12pt} \\
   \hline
   \small training set\textcolor{white}{\_} & \small {\hspace{1px}}3000{\hspace{1px}} & \small {\hspace{1px}}6000{\hspace{1px}} & \small {\hspace{1px}}9000{\hspace{1px}} & \small {\hspace{1px}}12000{\hspace{1px}} & \small {\hspace{1px}}15000{\hspace{1px}} & \small {\hspace{1px}}18000{\hspace{1px}} & \small {\hspace{1px}}21000{\hspace{1px}} \\
   \hline
   \vspace{-12pt} \\
   \small TF & \small .745 & \small .747 & \small .754 & \small .757 & \small .757 & \small .757 & \small .756 \\
   \small TF-IRF & \small \textbf{.800} & \small .808 & \small .813 & \small \textbf{.817} & \small .816 & \small .817 & \small .816 \\
   \small TF-IBF & \small \textbf{.800} & \small \textbf{.809} & \small \textbf{.814} & \small \textbf{.817} & \small \textbf{.817} & \small \textbf{.818} & \small \textbf{.818} \\
   \small TF-IUF & \small .797 & \small .805 & \small .810 & \small .814 & \small .813 & \small .814 & \small .814 \\
   \vspace{-12pt} \\
   \hline
   \vspace{-12pt} \\
   \multicolumn{8}{ c }{\small \textbf{GoodReads - LCC}} \\
   \vspace{-12pt} \\
   \hline
   \small training set\textcolor{white}{\_} & \small {\hspace{1px}}3000{\hspace{1px}} & \small {\hspace{1px}}6000{\hspace{1px}} & \small {\hspace{1px}}9000{\hspace{1px}} & \small {\hspace{1px}}12000{\hspace{1px}} & \small {\hspace{1px}}15000{\hspace{1px}} & \small {\hspace{1px}}18000{\hspace{1px}} & \small {\hspace{1px}}21000{\hspace{1px}} \\
   \hline
   \vspace{-12pt} \\
   \small TF & \small .725 & \small .731 & \small .737 & \small .738 & \small .734 & \small .731 & \small .743 \\
   \small TF-IRF & \small \textbf{.781} & \small \textbf{.792} & \small \textbf{.797} & \small .801 & \small \textbf{.802} & \small .799 & \small .804 \\
   \small TF-IBF & \small \textbf{.781} & \small \textbf{.792} & \small \textbf{.797} & \small \textbf{.803} & \small \textbf{.802} & \small \textbf{.800} & \small \textbf{.805} \\
   \small TF-IUF & \small .776 & \small .788 & \small .792 & \small .797 & \small .797 & \small .794 & \small .800 \\
   \vspace{-12pt} \\
   \hline
  \end{tabular}
 \end{center}
 \caption{Accuracy results of classification using weighting functions.}
 \label{tab:svm-idf}
\end{table}

Table \ref{tab:svm-idf} shows the results of using tag weighting functions. For Delicious, the outperformance of TF shows that the use of weighting functions is not useful in this case. Going further into the analysis of the performance of representations relying on weighting functions, the results show that IUF gets the best results among them, followed by IBF, and then IRF. We believe that resource-based tag suggestions, as occurs on Delicious, are not helpful to this end. It makes the top tags become even more popular and it alters the natural distribution of tags. Thus, such a forced distribution of tags produces weights that score lower performances. Moreover, the fact that IUF is the best weighting function in this case, shows the importance of users who make their own choices instead of relying on suggestions. That is, users who differ from suggestion-based annotations give rise to higher weights for their seldom tags, which performs better than IRF and IBF.

In the case of LibraryThing, all the weighting functions are clearly superior to TF, since the former always outperform the latter. This shows that the studied inverse weighting functions can be really useful for folksonomies created in the absence of suggestions. Tag weighting functions have successfully set suitable weights towards a definition of the representativity of tags in this case, in contrast to Delicious. Among the tag weighting functions, all of them perform similarly, and no clear outperformances can be seen in these results. However, IBF seems to provide slightly better results than the other two approaches, followed by IRF. IUF is the worst function in this case, suggesting that the number of users choosing each tag is not the most relevant feature when no suggestions are given.

On GoodReads, tag weighting functions also clearly outperform the sole use of TF. As on LibraryThing, IBF performs the best among the weighting functions, followed by IRF, and then IUF. Even though there are suggestions on GoodReads too, they rely on users' tags, and thereby these suggestions can only be applied to different resources. This shows that the effect of personomy-based suggestions is much smaller, and it affects to a lower extent or does not almost affect the distrubution of tags, because suggestions do not spread to users.

Summarizing, results show that the studied inverse tag weighting functions can be really useful for determining the representativity of tags within the collection. However, folksonomies can suffer from resource-based tag suggestions, transforming the structure and distributions of folksonomies. This transformation can even be harmful for the definition of the tag weighting functions, and can bring about worse performance results than simply relying on TF, as happened on Delicious. Otherwise, in the absence of resource-based tag suggestions, the use of tag weighting functions improves the performance.

Comparing the results scored by tag weighting functions, it can be seen that IBF is always slightly better than IRF. The former is more detailed than the latter, because it considers the exact number of appearances of the tag besides the number of resources it appears in. Actually, IBF is the best approach for both LibraryThing and GoodReads, where there are no suggestions, or suggestions rely on user's personomy. When these suggestions rely on tags annotated earlier to the resource, as on Delicious, IUF performs better than the other two weighting functions, showing the relevance of users' ability to dismiss suggestions. However, even IUF is unable to outperform TF in this case.

\section{Conclusions}
\label{conclusions}

This work complements earlier research on tag-based resource classification by exploring the effect of tag distributions on systems with different settings. Adapting a weighting scheme such as TF-IDF helps better understand these distributions. To the best of our knowledge, this is the first research work exploiting tags from different systems for a resource classification task. The study on three large-scale datasets from systems with different settings has given rise to better understanding the classification performance on each system. We have performed a thorough analysis of tag distributions on the datasets, and have analyzed the effect of those distributions as to the classification of resources using tags.

Among the settings, tag suggestions have shown to influence the structure of folksonomies greatly. Users tend to choose among suggested tags rather than providing their own tags, producing different folksonomies where a few tags stand out as compared to the rest. Systems with resource-based suggestions produce different tag distributions with respect to the absence of suggestions. The use of an IDF-like scheme shows that it cannot help determine the relevance of tags in such cases, where TF yields better results. However, we have found that tag weighting functions clearly outperform the TF approach when resource-based tag suggestions are disabled, i.e., on LibraryThing and GoodReads. Our findings are relevant for scientists studying tag-based resource classification, user behavior in social networks, the structure of folksonomies and tag distributions, as well as for developers of tagging systems in search of the appropriate setting.

Future work includes studying the suitability of inverse tag weighting functions for other information management tasks, and further analyzing suggestion biases like Delicious, in search of a weighting function to fit them.

\bibliographystyle{splncs03}
\bibliography{twf-ksem2011}

\end{document}